\documentclass[dvips]{elsart}
\usepackage{graphicx}
\begin{document}
\begin{frontmatter}
\title{Viscosity models for silicate melts}
\maketitle
\author{Jesper deClaville Christiansen}
\address{Center for Nanotechnology, Aalborg University
Fibigerstraede 16, DK-9220 Aalborg East, Denmark
E-mail: i9ictus@iprod.auc.dk, Phone: +45 9635 8970}
\maketitle
\begin{abstract}
Rheology of silicates (melts or super cooled as glasses) is one of the most important fields to mankind, 
as geological activity largely influences life. In this paper, unifying empirical models for the Non-Newtonian 
shear viscosity and the extensional viscosity are proposed. The models have only one parameter, the 
zero shear viscosity (the Newtonian plateau), and are valid for a vide range of chemical compositions
\end{abstract}
\end{frontmatter}
\section{Introduction}
The field of geology attracts scientists from many areas. This scientific field has for 
obvious reasons a great importance to mankind. In geology, the study of silicate melts attracts the main
focus because in nearly all igneous processes silicate melts does play a role. The importance of the
field attract many researchers in the search for general models and physical origins to the behaviour of 
silicate melts. Due to the chemical complexity of silicates occurring in nature the area 
of rheology of silicates is not easiest field one can choose. Amongst the many studies 
\cite{dingwella,dingwellb,bagdassarov,bagdassarovb,bagdassarovc,dingwellc,dingwelld,dingwelle,bottinga,bottingab,stebbins,simmons,simmonsb} 
can be mentioned. The challenges in rheometry can be seen, when
considering the need for temperature spans of more than 2000 C, and viscosities ranging from 
$10^{40}$ Pas \cite{walters} at room temperature to $10^{12}$ Pas at $T_{g}$ and $10^{-4}$ Pas at supercritical
conditions. Finally the melts are chemically very aggressive in the high temperature range. 
Rheometrical methods that have been used in the study of silicates are nearly all that we know: 
Concentric cylinder, cone/plate, "solid" bar oscillatory rheometry, ball indentation, fiber elongation,
biaxial compression, ultrasonics and capillary rheometry. Many of these methods does not provide the accuracy 
that we normally are used to when doing rheometry under more pleasant environmental conditions. In this study
an analysis have been made on experimental oscillatory data from \cite{GLAFO1,GLAFO2}  and on extensional data from \cite{dingwella,dingwellb}. 

\section{Analysis of experiments}
\begin{figure}[htb]
\centering
\includegraphics[width=65mm]{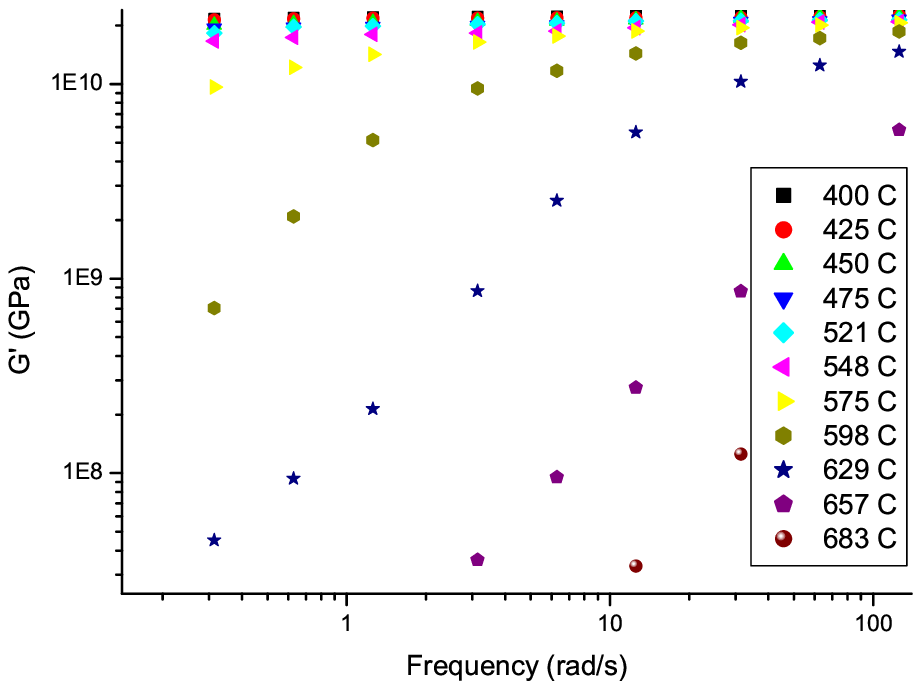} \includegraphics[width=65mm]{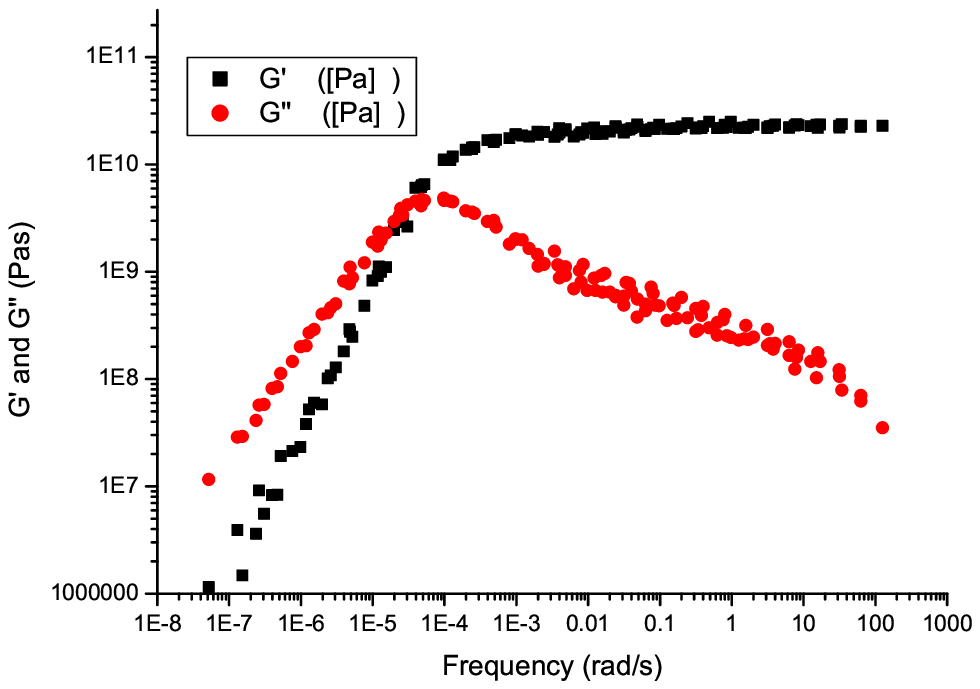}
\caption{Raw oscillatory data (673 K - 956 K) for the glass fibre composition and the master curve (902 K)
obtained from time-temperatures superposition using IRIS \cite{IRIS}.}
\label{fig:1}
\end{figure}

The raw data for the following analysis work was taken from \cite{GLAFO1} and \cite{GLAFO2}. In the first reference, 
N. Bagdassarov made oscillatory experiments for GLAFO, The Swedish Glass Research Institute, on an E-glass 
and a Rock wool composition, both below and above $T_{g}$. In the second reference, 9 different chemical compositions 
were measured by GLAFO on a Bohlin oscillatory rheometer, in both cases a "solid" bar type of instrument was used.
An example of the raw data for 11 different temperatures for a glass fibre composition can 
be seen on Figure~\ref{fig:1}. On the right hand side of Figure~\ref{fig:1}, the data have been shifted to a master 
curve at 902 K, using IRIS \cite{IRIS}. 

\begin{figure}[htb]
\centering
\includegraphics[width=80mm]{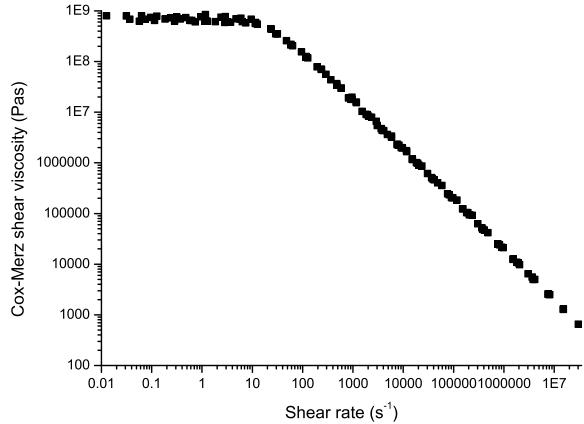}
\caption{The Cox-Merz shear viscosity as a function of shear rate, E-glass (902 K).}
\label{fig:2}
\end{figure}

The data falls quite nicely on a single master curve. Now it is possible to 
apply the Cox-Merz approximation \cite{walters} and obtain Figure~\ref{fig:2}. The Non-Newtonian onset corresponds 
to a Deborah number of approximately 0,7 when G is assumed constant 25 GPa no matter temperature and composition. 
The latter approximation is acceptable, as the viscosity can vary several decades, whereas G only varies a few 
percent, and remains in the same range even with different chemical compositions. When making this exercise for 
all the different chemical compositions and temperatures and plotting the Cox-Merz 
shear viscosity, a very interesting behaviour is observed: The Deborah number for the Non-Newtonian onset 
remains constant around 0,7. This observation agrees well 
with the discussion of Webb and Dingwell in \cite{stebbins}.\\

Similar observations for the extensional viscosity were found by \cite{dingwella}. From the data from \cite{dingwella}, 
the onset can be calculated to occur at a Deborah number approximately two decades lower than what was 
observed in this study. This difference is consistent with what one would expect, as an extensional flow is 
known to be much stronger in aligning structure than shear, thus the onset is lower in an extensional flow. 
What is really surprising in the study of the extensional properties in \cite{dingwella,dingwellb} is the fact, 
that silicates also in extension show an extensional thinning behaviour. But this can be explained from 
the extremely small size of structures in a silicate melt compared with e.g. polymers.\\ 

\begin{figure}[htb]
\centering
\includegraphics[width=90mm]{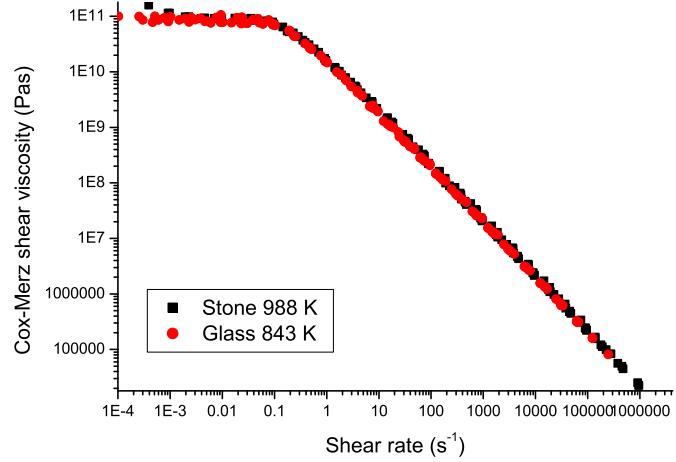}
\caption{The Cox-Merz shear viscosity for an E-glass (843 K) and a rock wool composition (988 K), with temperatures 
adjusted to provide the same zero-shear viscosity.}
\label{fig:3}
\end{figure}

In the current study, it was observed  from the shear viscosity plots that the slope of the 
graph in the shear thinning region is constant for all compositions. Thus it is possible to 
merge all shear viscosity curves of the different silicate melts to a single curve for the same 
Newtonian plateau. This is shown on Figure~\ref{fig:3} for two compositions, the E-glas and the Rock wool, 
(only two in order to make the plot easy to read). In \cite{dingwella} the same was made for the extensional case, 
and in \cite{lundjensen} for a single composition but different temperatures. 
A natural question that can be raised is the origin for this simplicity, when we do have a non-exponential
nature of the relaxation (a pluralis of relaxation times). 
A good discussion have been made by Webb and Dingwell in \cite{stebbins}.

\section{A Non-Newtonian shear viscosity model for silicates}
With the knowledge from Figure~\ref{fig:3}, it is now possible to propose a shear viscosity 
relation that fits all the chemical 
compositions used in this study. A viscosity model is just curve fit; the choice of a good mathematical model to 
fit experimental data. Many models are all ready available that represents various degrees of 
complexity depending on the number of parameters. For this study, a simple Carreau type of model was chosen.

\begin{equation}
\eta({\dot \gamma}) = \frac{A_{0}}{\Big(1 + (A_{1}{\dot \gamma})^{2}\Big)^{A_{2}}}
\end{equation}

$A_{0}$ should be the Newtonian shear viscosity plateau at low shear rates. 
$A_{2}$ affects the slope of the curve in the Non-Newtonian region and 
was found from fitting experiments on the data sets to be 0,48. $A_{1}$ is the parameter that adjusts the 
Non-Newtonian onset. 
We know from the observations in this study , that the onset approximately occur at a Deborah number of 0,7. 
With this and the definition of the Deborah number \cite{walters}, an onset shear rate can be defined: 

\begin{equation}
{\dot \gamma}_{onset} = \frac{DeG_{\infty}}{\eta_{0}} \Rightarrow
\end{equation}

\begin{equation}
{\dot \gamma}_{onset} = \frac{0.7G_{\infty}}{\eta_{0}} 
\end{equation}

$A_{1}$ is found to be the reciprocal to the onset shear rate. We now have a suggestion to a simple model for 
the shear viscosity for silicate melts

\begin{equation}
\eta({\dot \gamma}) =\frac{\eta_{0}}{\Big(1 + ( \frac{\eta_{0}}{0.7G_{\infty}}{\dot \gamma})^{2}\Big)^{0.48}}
\end{equation}

\begin{figure}[htb]
\centering
\includegraphics[width=65mm]{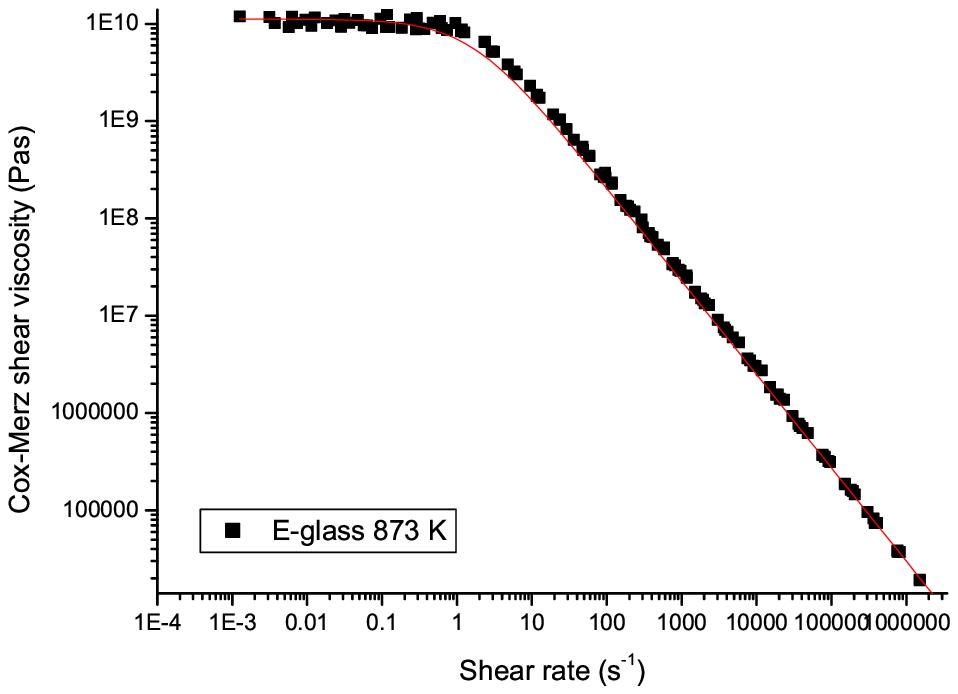} \includegraphics[width=65mm]{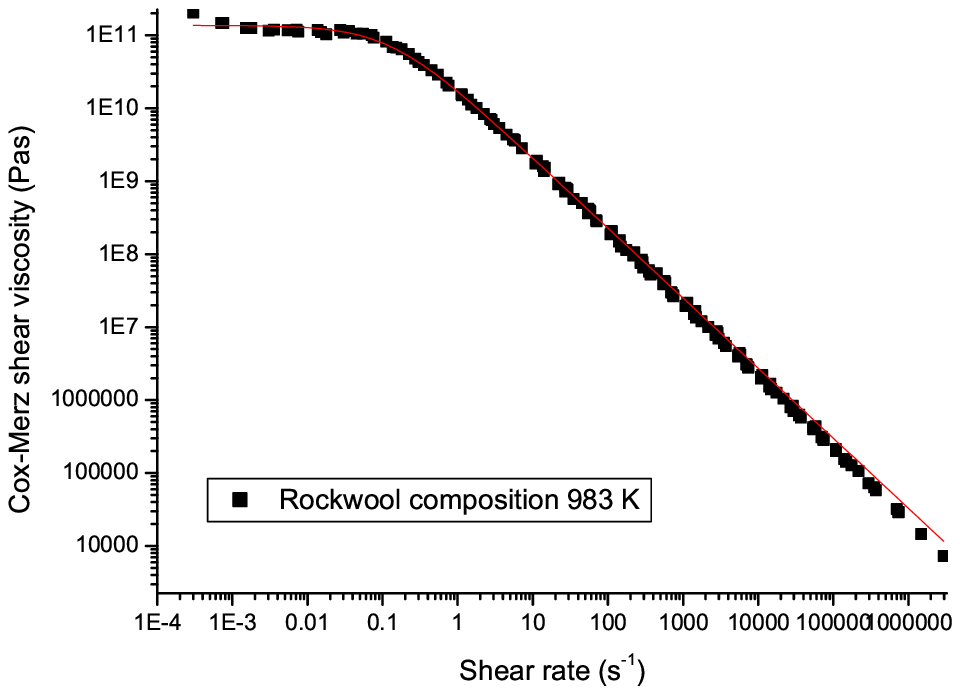} \\
\caption{The figure shows the Cox-Merx shear viscosity (points) for two different chemical compositions 
together with the prediction of the shear viscosity model proposed in this study. Analysis based on \cite{GLAFO1}}
\label{fig:4}
\end{figure}

\begin{figure}[htb]
\centering
\includegraphics[width=65mm]{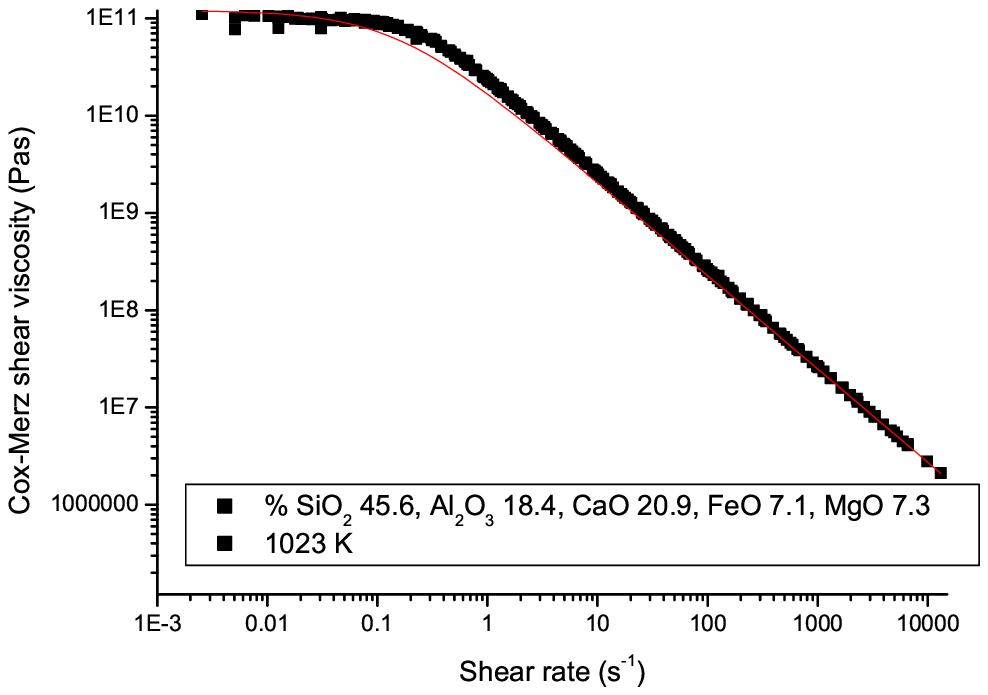} \includegraphics[width=65mm]{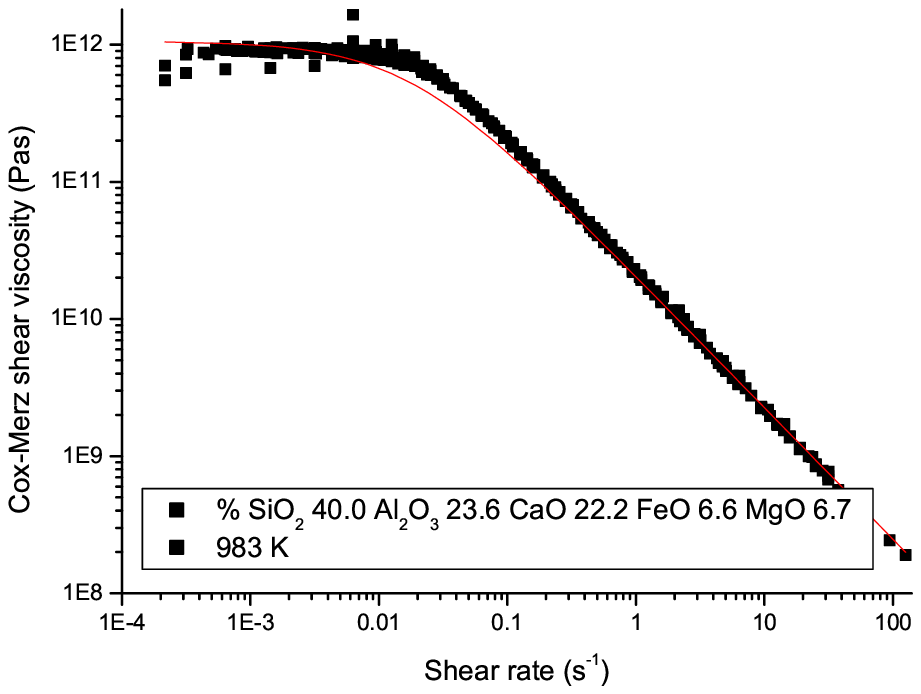} \\
\includegraphics[width=65mm]{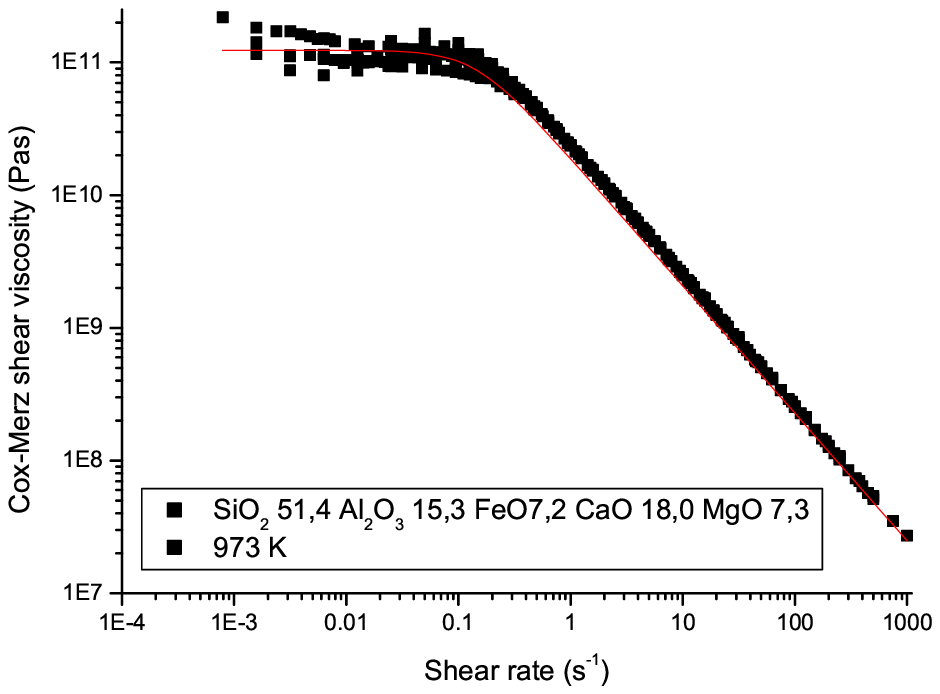} \includegraphics[width=65mm]{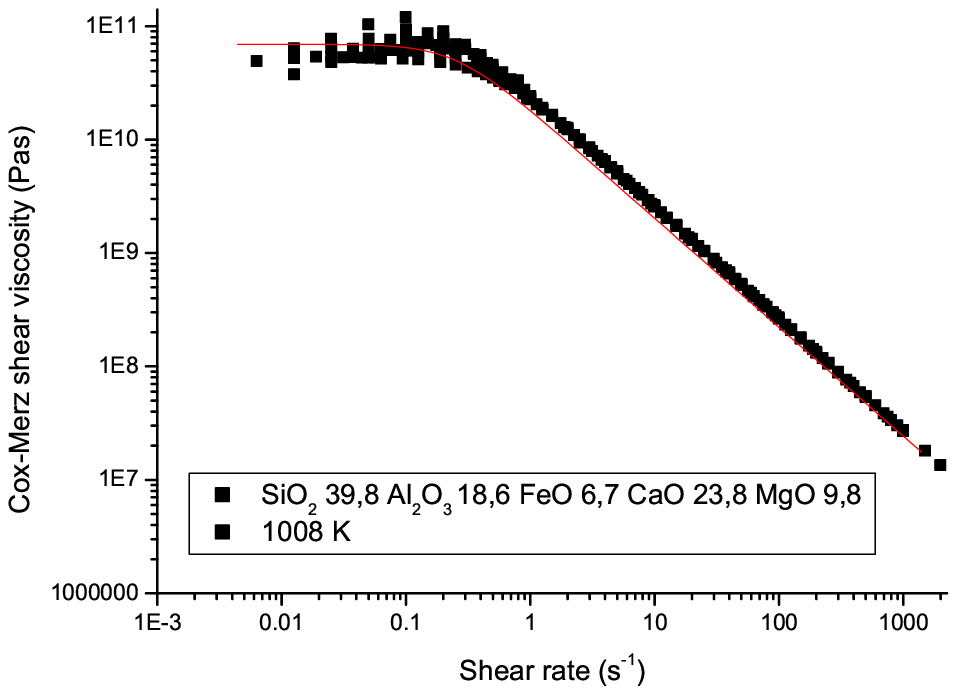}
\caption{The figure shows the Cox-Merz shear viscosity (points) for four different chemical compositions 
together with the prediction of the shear viscosity model proposed in this study. Analysis based on \cite{GLAFO2}}
\label{fig:5}
\end{figure}

Figure~\ref{fig:4} and Figure~\ref{fig:5}, show the comparison of experimental data and 
the prediction of this model on six different chemical compositions. 

The quality of the fits on Figures~\ref{fig:4} and Figures~\ref{fig:5} depends on the 
quality of the oscillatory data as some datasets show mores scatter than others. The slight deviation af the 
data and the curve is due to this scatter in the lower shear region for some measurements, 
as the curve was drawn in Origin, using a Non-linear curve fit to 
provide the "best" fit. All together the model (4) show a fair fit to the data.

\section{Extensional viscosity}
Knowing, that the zero-extensional viscosity is three times the zero-shear viscosity and having the data from 
the article of \cite{dingwella,dingwellb} a releation for the extensional viscosity can be proposed.

\begin{equation}
\eta_{E}({\dot \epsilon}) =\frac{3\eta_{0}}{ \Big(1 + ( \frac{3\eta_{0}}{De3G_{\infty}}{\dot \epsilon})^{2}\Big)^{A_{2}}} \Rightarrow
\end{equation}

\begin{equation}
\eta_{E}({\dot \epsilon}) =\frac{3\eta_{0}}{\Big(1 + ( \frac{\eta_{0}}{DeG_{\infty}}{\dot \epsilon})^{2}\Big)^{A_{2}}}
\end{equation}

Here in equation (5) the three times the shear viscosity can be replaced by measured data from an extensional 
experiment and 3G with Youngs Modulus E – if extensional data are available. 
From \cite{dingwella,dingwellb}, the De number for the onset can be found to approximately 0,003. 
The lower De number for the onset corresponds well to the fact that extension is a structural stronger 
aligning flow than shear. Thus a first suggestion for an equation would be:

\begin{equation}
\eta_{E}({\dot \epsilon}) =\frac{3\eta_{0}}{\Big(1 + ( \frac{\eta_{0}}{0.003G_{\infty}}{\dot \epsilon})^{2}\Big)^{0.48}}
\end{equation}

\begin{figure}[htb]
\centering
\includegraphics[width=65mm]{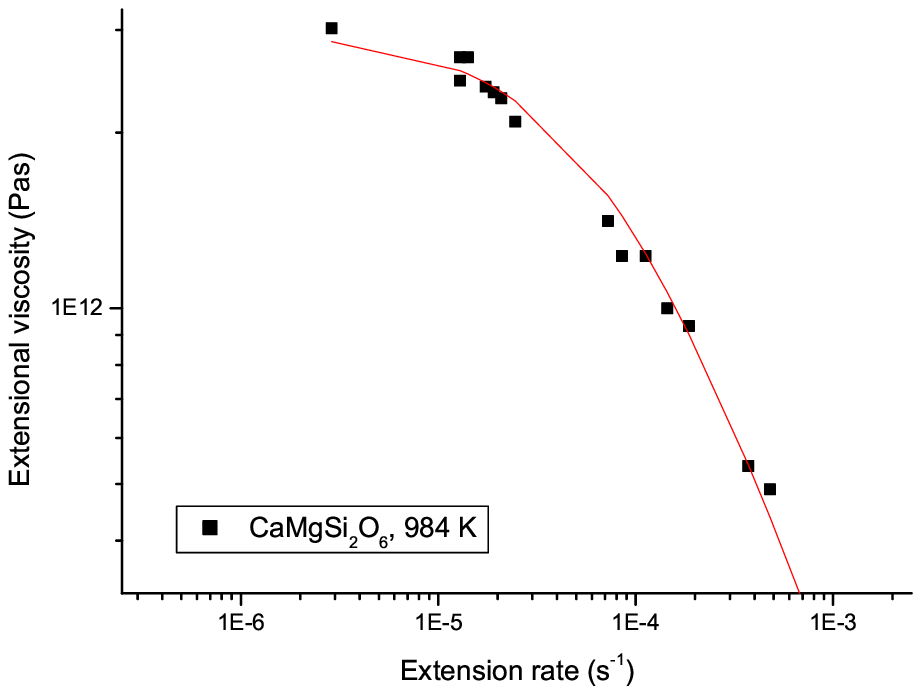} \includegraphics[width=65mm]{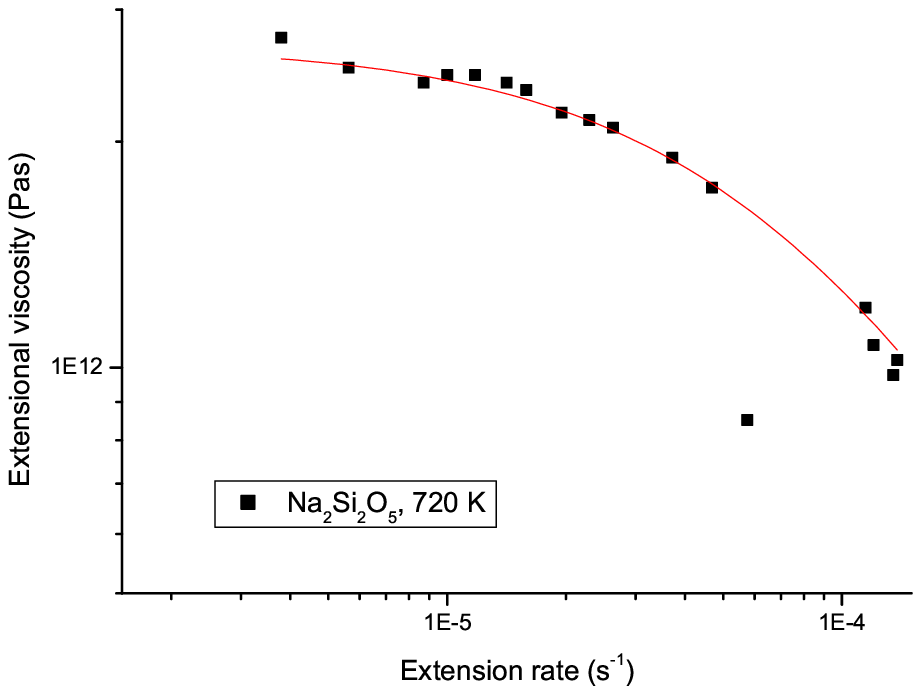}
\caption{Comparison between data from \cite{dingwella,dingwellb} and the model. The line represents 
the best fit of the model to the data (using Origin with only unknown $\eta_{0}$).}
\label{fig:6}
\end{figure}

\begin{figure}[htb]
\centering
\includegraphics[width=65mm]{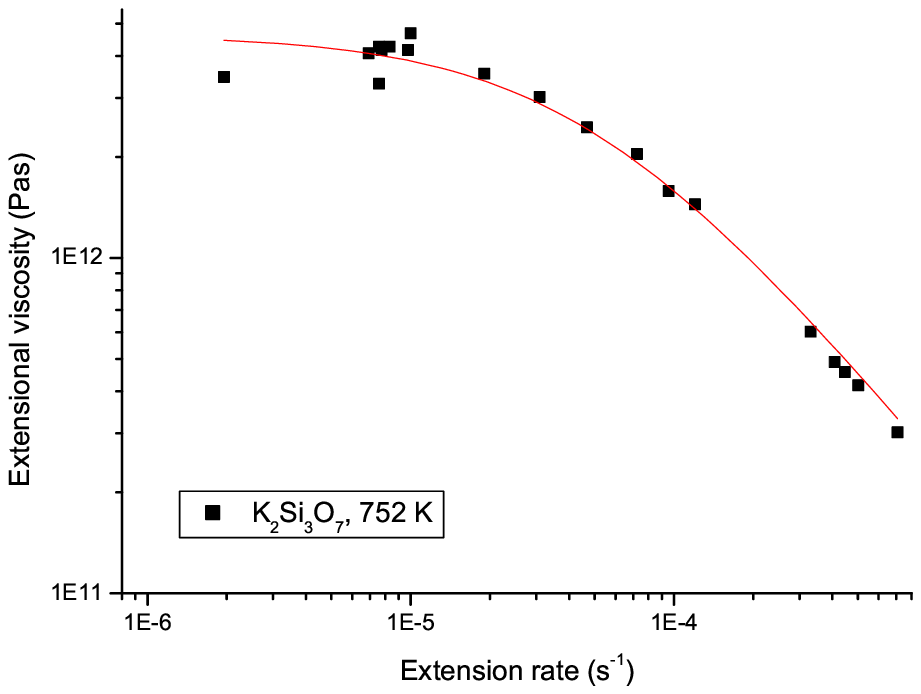} \includegraphics[width=65mm]{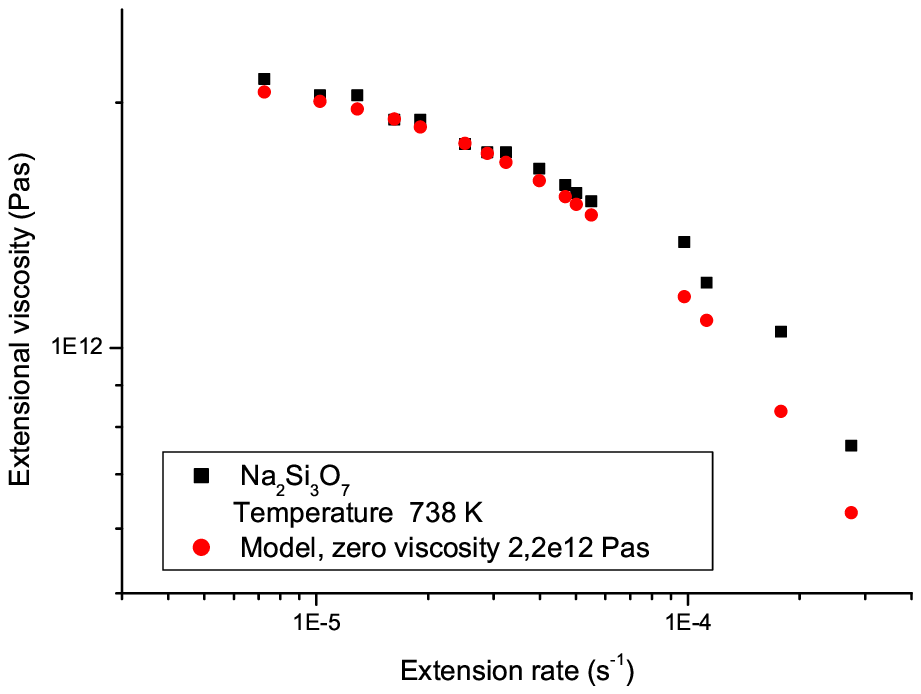}
\caption{Comparison between data from \cite{dingwella,dingwellb} and the model. The line (left figure) represents 
the best fit of the model to the data (using Origin). The points (right figure) were 
obtained by manually setting a value for
the zero shear viscosity.}
\label{fig:7}
\end{figure}

\begin{figure}[htb]
\centering
\includegraphics[width=65mm]{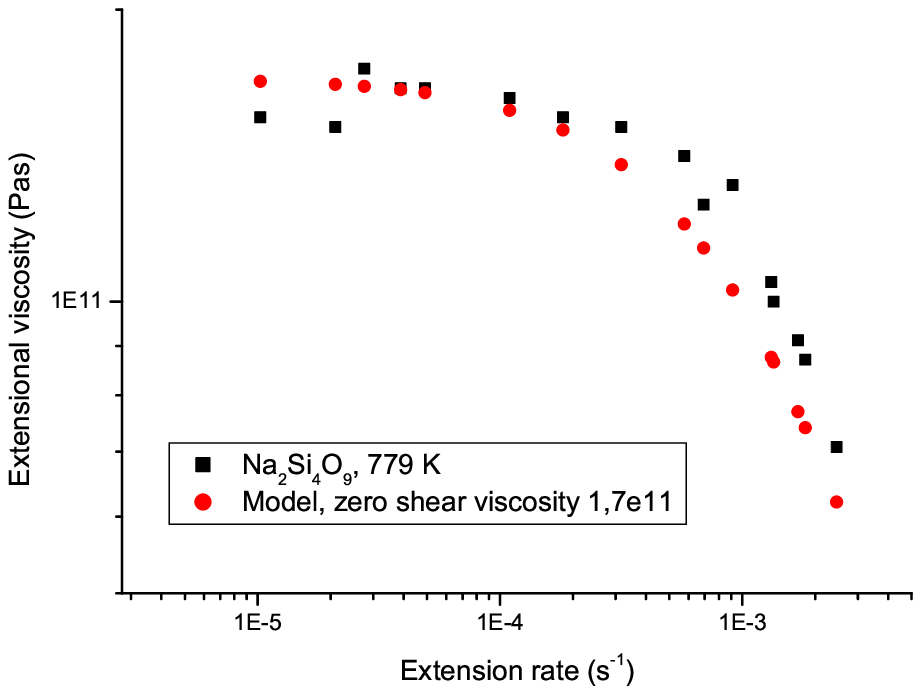} \includegraphics[width=65mm]{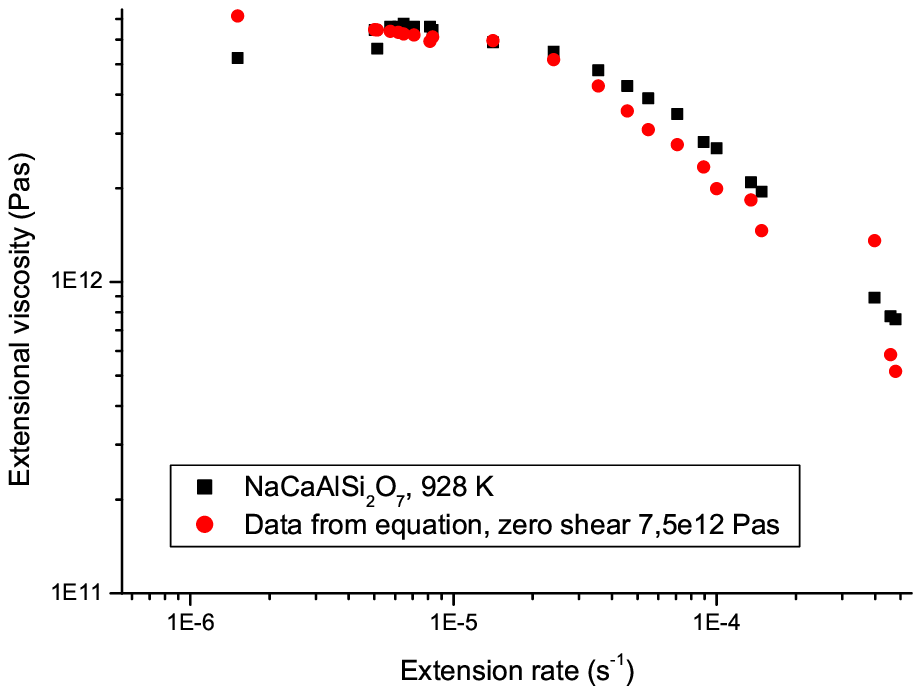}
\caption{Comparison between data from \cite{dingwella,dingwellb} and the model. The points were obtained by manually setting a value for
the zero shear viscosity.}
\label{fig:8}
\end{figure}

This equation have been tested on the data from \cite{dingwella,dingwellb}. 
Results can be seen in Figure~\ref{fig:6},\ref{fig:7} and \ref{fig:8}. 
The suggested model can be seen to represent a fair fit to
the data. The model is almost identical to the model of \cite{simmons,simmonsb}, 
but it has the advantage, 
that it can be adjusted through the De number. 

\section{Discussion and conclusions}
Oscillatory data from two different laboratories on 11 different chemical compositions have been analysed. 
Time-temperature superposition have been performed on all dataset, and the Cox-Merz shear viscosity plotted. 

It is observed, that the onset to Non-Newtonian behaviour for all chemical compositions in shear 
occur in the order of a 
Deborah number of 0,7. All curves can be shifted to a master curve, when scaled to the same Newtonian plateau 
viscosity. Data was fitted using the Carreau model, and it was observed, that the parameters $A_{1}$ could be found 
as the reciprocal of the onset shear rate. The onset shear rate can be calculated from the Deborah number. 

$A_{2}$ was found by fitting experiments to be approximately 0,48. $A_{0}$ is simply the zero-shear 
viscosity (Newtonian plateau). 

Thus an empiric expression has been proposed, that allows an estimate of the Non-Newtonian shear viscosity curve, 
and where the only variable is the Newtonian viscosity; an easily found parameter that can be estimated from 
the viscosity-temperature function that are available in 
literature \cite{solvang} for many different chemical compositions. 
The model is in extension quite similar to the work of Simmons \cite{simmons,simmonsb}, but has the adjustable 
feature of a De number. 

The applicability of such relations as proposed here are potentially very big. Geological processes involve 
variations in chemical composition, and phenomena like melt fracture occur in the Non Newtonian range.  

The results reported here and the model proposed agree well with the observations of \cite{dingwella,dingwellb}. 
In fact it was possible to make a fair fit to the data from this work just modifying the 
shear equation to an extensional viscosity model with a De=0,003 to compensate for a stronger aligning flow. 
The other modifications made was straight forward: E=3G and the zero-extensional viscosity equals 
three times the zero-shear viscosity. The model was demonstrated to fit extensional data from 
\cite{dingwella,dingwellb} very well.

The simple relation between \cite{dingwella,dingwellb} and shear is nice, as it strengthens 
the conclusion of the applicability of the Cox-Merz approximation after the Non-Newtonian onset
and simplifies the models. 

Though robust to chemical compositions future work is needed in order to verify the limits of this model 
regarding compositions. Adjustments to suggested values of Deborah number for the onset and the parameter 
$A_{2}$ would also be welcome. The Carreau model was chosen as a fit function, but others fit functions 
can be tested. 
The chosen value of G of 25 GPa is a minor approximation, 
as G varies slightly with composition, temperature and pressure. 

\section{Acknowledgements}
The author would greatly like to acknowledge GLAFO, the Swedish Glass Research Institute, for making data 
available for the analysis. Also the author would like to thank Dr Erik Appel Jensen, Aalborg University 
and Dr. S\o ren Lund Jensen, Rockwool International for good discussions and 
Rockwool International A/S for financial support.

\end{document}